# Policy options for digital infrastructure strategies: A simulation model for broadband universal service in Africa

Edward Oughton, George Mason University

## Abstract

Internet access is essential for economic development and helping to deliver the Sustainable Development Goals, especially as even basic broadband can revolutionize available economic opportunities. Yet, more than one billion people still live without internet access. Governments must make strategic choices to connect these citizens, but currently have few independent, transparent and scientifically reproducible assessments to rely on. This paper develops open-source software to test broadband universal service strategies which meet the 10 Mbps target being considered by the UN Broadband Commission. The private and government costs of different infrastructure decisions are quantified in six East and West African countries (Côte D'Ivoire, Mali, Senegal, Kenya, Tanzania and Uganda). The results provide strong evidence that 'leapfrogging' straight to 4G in unconnected areas is the least-cost option for providing broadband universal service, with savings between 13-51% over 3G. The results also demonstrate how the extraction of spectrum and tax revenues in unviable markets provide no net benefit, as for every $1 taken in revenue, a $1 infrastructure subsidy is required from government to achieve broadband universal service. Importantly, the use of a Shared Rural Network in unviable locations provides impressive cost savings (up to 78%), while retaining the benefits of dynamic infrastructure competition in viable urban and suburban areas. This paper provides evidence to design national and international policies aimed at broadband universal service.

## Key words

## 1. Introduction

Broadband has long been recognized as critical for helping to deliver the Sustainable Development Goals (SDGs), and enable digitally-led development (Mansell, 2001, 1999; Mansell and Wehn, 1998). Indeed, governments are increasingly treating digital infrastructure on par with energy or water access, given its importance for economic development (Chen et al., 2020; Czernich et al., 2011; Koutroumpis, 2009; Oughton et al., 2015; Röller and Waverman, 2001). Even basic access transforms the opportunities available to citizens (Aker, 2011; Aker and Mbiti, 2010; Suri and Jack, 2016).

Over 3.2 billion people are connected globally to the internet via cellular, leaving a significant digital divide. Approximately 2.6 billion people live within cell coverage but without a handset ('the usage gap'), while the remaining 1.6 billion people live without coverage ('the coverage gap') (GSMA, 2017). Currently the ITU has set a range of targets to be achieved by 2025 for internet access globally (International Telecommunication Union, 2019) including bringing 75% of the global population online by 2025. The UN Broadband Commission has been exploring the implications of a 10 Mbps per user universal target, but hitherto there has been scant analysis of the level of investment involved.

Understanding the economics of internet infrastructure is essential (Claffy and Clark, 2019). Despite high-level policy ambitions, how universal broadband service should be delivered globally is still unknown. Whereas universal service has been a cornerstone of regulatory policy in networked industries for many decades in frontier economies (Cremer et al., 2001), greater emphasis is now being placed on this concept in emerging economies for broadband. In high income countries, universal service has enabled equality of access while shifting monopoly industries towards market-based competition. Yet, in emerging economies most assets need to be built from scratch, requiring considerable investment. Two research questions are therefore identified for investigation including:

1. Which technologies should governments encourage to enable broadband universal service?
2. What level of infrastructure sharing should governments encourage to help deliver broadband universal service?

Having outlined these research questions, a literature review is now undertaken before a suitable method is described in Section 3. The assessment results will then be presented in Section 4, with the ramifications discussed and paper conclusions given in Section 5.

## 2. Literature review

Universal service is defined as an operator providing a basic level of service to all potential users at an affordable rate. There are two main types of broadband technologies which are potential candidates, including fixed access (via a copper, coaxial or fiber cable), or wireless access (via cellular, Wi-Fi or satellite). However, even within these segments there are competing options, such as the competition between cellular versus Wi-Fi for wireless broadband connectivity (Oughton et al., 2020). Increasingly a hybrid approach which blends fixed and wireless is becoming common, such as Fixed Wireless Access (Abozariba et al., 2019), in an attempt to provide better service at lower cost. Each technology exhibits a different cost supply curve, making it more competitive in different deployment situations (Anusha et al., 2017), depending on the necessary capacity and coverage required in a local area. For example, in dense urban areas where traffic demand is very high, fixed fiber is much more economic than using wireless methods. In contrast, wireless access is much more economic in low density areas where there are fewer users, spread out over a wide area (Hameed et al., 2018; Lertsinsrubtavee et al., 2018), particularly as the initial capital investment can be lower. A variety of new technologies have been proposed for helping to fill coverage gaps in rural and remote areas. These range from incremental extensions of existing technologies, such as larger cells or using TV whitespaces (Khalil et al., 2017), to much more radical developments, ranging from Unmanned Aerial Vehicles (Chiaraviglio et al., 2017; Jiménez et al., 2018), to deployment of mass produced Low Earth Orbit satellite constellations (Saeed et al., 2020).

Backhaul connectivity remains one of the key challenges for serving remote locations, as new local access technologies need to be able to transport data to and from servers elsewhere in the internet. The costs of this data transportation are prohibitive in many locations, especially in mountainous areas where many line-of-sight connections could be required. Wireless backhaul links are generally preferred for terrestrial deployments in hard-to-reach areas. However, the civil engineering costs of erecting towers with line-of-sight paths can be high, particularly when deploying in challenging environments. Equipment may need to be transported using an unsealed road and carried by hand uphill to proposed sites. Water may not be available to mix concrete, meaning it needs to be carried by hand. Vegetation may need to be manually reduced to enable a structure to be erected, and any tower needs to clear the tree line to provide adequate propagation conditions.

In locations where the costs of delivery exceed the potential revenue a Mobile Network Operator (MNO) could achieve, market failure occurs leading to a lack of infrastructure investment. In such a situation, appropriate measures need to be taken to enable viability, and one such change can be the sharing of infrastructure assets, to help reduce cost (Oughton et al., 2018). Hence, infrastructure sharing strategies are particularly pertinent for MNOs in emerging markets (Meddour et al., 2011). There has been less of a need to share infrastructure in earlier generations, such as during the 2G era, as MNOs experienced increasing revenues and benefited from very large cell areas. Currently however, revenues are either static or declining in many global telecommunication markets. Four main types of infrastructure sharing options are illustrated in Figure 1.

Figure 1 Infrastructure sharing strategies (GSMA, 2019a)

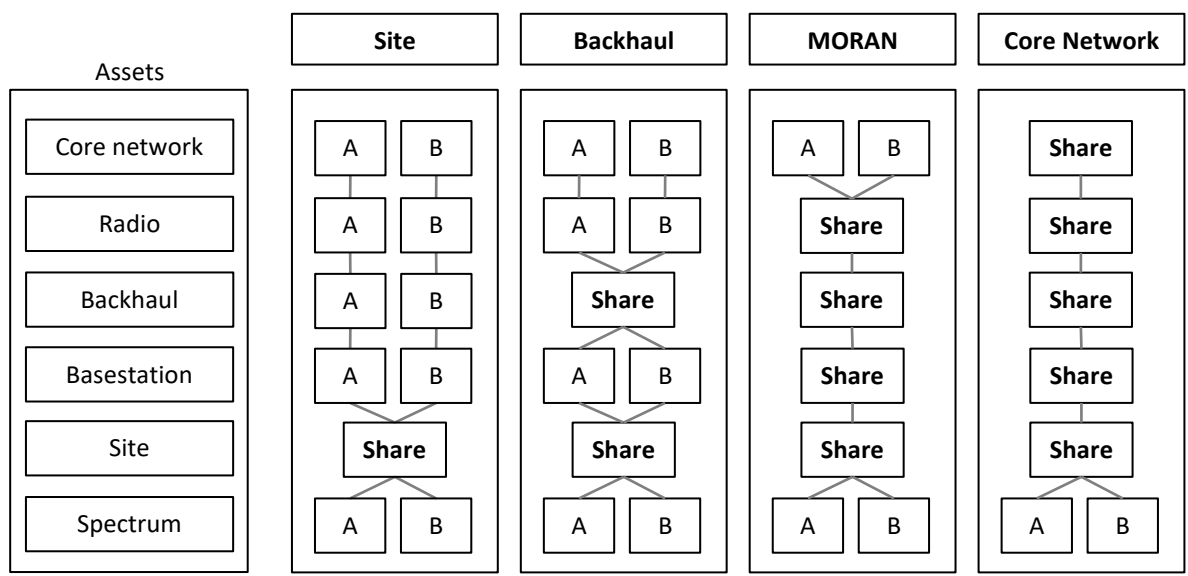

Sharing active equipment has a beneficial impact on lowering operational expenses such as energy consumption (Antonopoulos et al., 2015; Bousia et al., 2016), particularly in areas with low demand where assets are not close to their full capacity. Spectrum sharing strategies introduce efficiency benefits, such as coordinating interference and providing carrier aggregation, reducing the number of required sites, improving economic viability, essential because spectrum is often underutilized (Boulos et al., 2020; Frias et al., 2020; Gomez et al., 2019; Jurdi et al., 2018; Peha, 2009). Moreover, network resource sharing helps to expedite the time taken to achieve viability for greenfield infrastructure, which is essential for rural and remote areas (Mamushiane et al., 2018).

However, a key caveat is that infrastructure sharing benefits need to be traded-off against any potential negative impacts (Sanguanpuak et al., 2019). For example, infrastructure competition is known to produce positive consumer outcomes, therefore consolidation needs to be assessed in terms of how it affects dynamic competition (Wallsten, 2005, 2001; Yoo, 2017). One option is to utilize infrastructure sharing only in unviable locations (e.g. a Shared Rural Network), while preserving the benefits of competition in viable areas (e.g. urban and suburban). Evidence suggests that even in competitive markets where one operator could drive another out of business, infrastructure sharing is still beneficial to the dominant player (Andrews et al., 2017). A method will now be presented.

## 3. Method

Myriad high-level policy reports have attempted to quantify the costs of infrastructure delivery for connecting unconnected communities. The majority use spreadsheet methods to estimate the required investment, leaving substantial uncertainty embedded within the results which is rarely portrayed to policy makers. The method developed here takes a new approach by drawing on a range of analytics tool rarely utilized in telecom policy research, including remote sensing and least-cost network designs derived from infrastructure simulation. Figure 2 illustrates how these approaches are combined to produce demand and supply estimates to quantify broadband universal service strategies. The open-source codebase adheres to gold-standard software engineering methods (fully tested, fully documented) and is openly available from the Policy Options for Digital Infrastructure Strategies (PODIS) repository: https://github.com/edwardoughton/podis

Figure 2 Quantifying broadband universal service strategies using data analytics methods

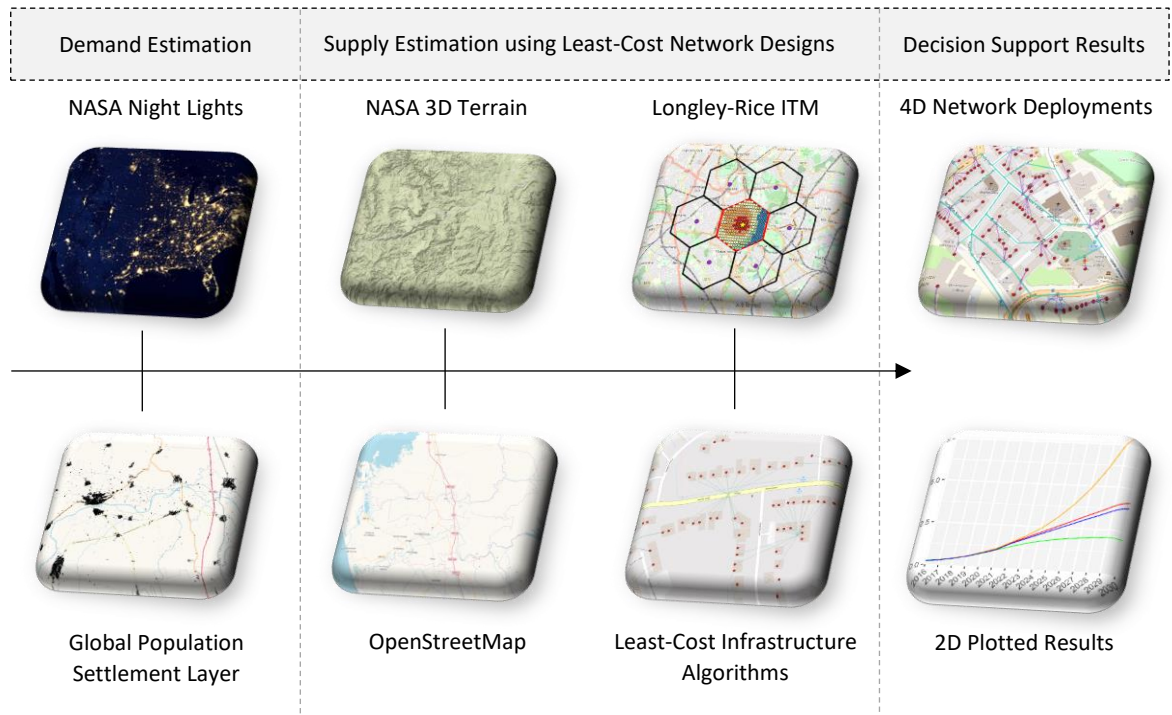

A scenario-based approach is used to assess the research questions, as is common in the literature to test 'what if' questions (Paltsev, 2017; Postma and Liebl, 2005; Swart et al., 2004) for infrastructure decisions (Hall et al., 2016a, 2016b; Thoung et al., 2016), including digital infrastructure roll-out (E.J. Oughton and Russell, 2020, 2020). This involves using a standard regulatory approach for making telecom policy decisions by modeling a representative '*Hypothetical Mobile Network Operator*', based on a Long-Run Incremental Cost modeling approach (Ofcom, 2018). The method is applied to assess six East and West African countries over the next decade (Côte D'Ivoire, Mali, Senegal, Kenya, Tanzania and Uganda).

## 3.1. Strategies

The research questions represent some of the main decisions governments currently face when designing policies for broadband universal service, ranging from which technologies to promote private operators to deploy, to the degree of infrastructure sharing desired. For decisions around the types of technologies that could be used, the focus is placed on a cellular approach, exploring whether 3G or 4G should be deployed in the Radio Access Network (RAN). Additionally, as many of the areas yet to receive a basic level of coverage are rural and often remote, the backhaul technology is a significant cost component (Ignacio et al., 2020). Wireless backhaul is likely to be cheaper, than a fixed fiber link, thanks to lower capital expenditure (capex), but the fiber can serve much higher traffic demand and has lower operational expenditure (opex) over the long-term.

Different types of infrastructure sharing are to be tested, reflecting the options identified in the literature review. In the baseline each MNO builds their own network to serve their market share with no sharing taking place. The other infrastructure sharing strategies include (i) passive site sharing (site compounds are shared), (ii) passive backhaul sharing (backhaul and site compounds are shared), (iii) active network sharing (Multi Operator Radio Access Network – MORAN), (iv) a shared rural network (using a MORAN plus a shared core network only in rural areas). Having outlined the main strategies, the demand and supply estimation steps are now articulated.

## 3.2. High-resolution spatial estimation of traffic and revenue

In the demand estimation module, the traffic demand ($Demand_i$) (Mbps km$^2$) for local smartphone users in the $i$th local region at time $t$ is estimated using data on the total population ($Pop_i$), cell phone penetration ($Pen_{it}$), smartphone penetration ($SP_{Pen_{it}}$) and a desired minimum per user capacity ($Cap_i$). The formula used to make this estimation is described in equation (1) and the maximum data demand for all years (2020-2030) is used to represent the peak traffic load.

$$Demand_i = \frac{Pop_i \cdot Pen_{it} \cdot SP_{Pen_{it}} \cdot Cap_i}{Market_{Share_i}} / OBF / Area_i \qquad (1)$$

The population is extracted from the WorldPop 2020 global raster layer (Stevens et al., 2015; Tatem, 2017), the cell phone penetration is treated as the GSMA unique number of mobile subscribers (GSMA, 2020), and the smartphone adoption rate is taken from the Research ICT Africa After Access

Survey (Research ICT Africa, 2018). The per user capacity level ($Cap_i$) is exogenously treated as a minimum universal broadband target of 10 Mbps per user, along with an exogenously stated market share ($Market\_Share_i$) depending on how many MNOs are present in the market being modeled. As not all users access the network simultaneously, an overbooking factor ($OBF$) of 20 is used as is standard in cellular dimensions (Holma and Toskala, 2011). Both the number of unique mobile subscribers and the smartphone adoption rate are forecast forward as a set of exogenous inputs for the simulation model, as illustrated with the scenario trends in Figure 3.

Figure 3 Unique mobile subscribers by country

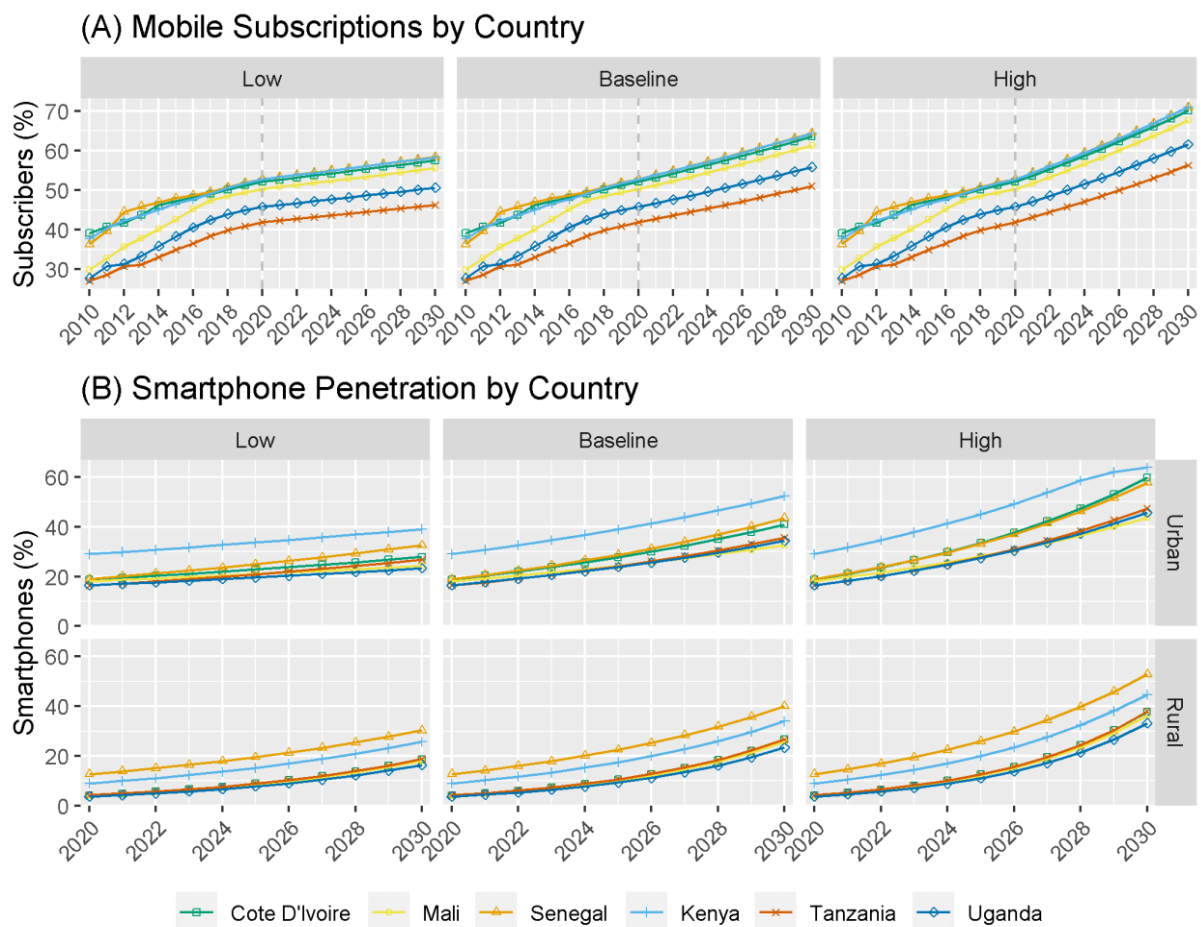

The revenue ($Revenue_i$) is estimated in a similar way for each region, except the exogenous per user capacity is substituted for the Average Revenue Per User ($ARPU_i$), as illustrated in equation (2).

$$Revenue_i = \frac{Pop_i \cdot Pen_{it} \cdot SP_{Pen_{it}} \cdot ARPU_i}{Market\_Share_i} \qquad (2)$$

The monthly Average Revenue Per User (ARPU) is estimated by remotely sensing nightlight luminosity from NASA VIIRS data, as a predictor of economic consumption in local areas, in order to allocate economic consumption tiers (Bruederle and Hodler, 2018; Henderson et al., 2012, 2011; Oughton and Mathur, 2020). Nightlight luminosity is measured in 'Digital Numbers' (DN) and the mean luminosity is used to allocate high, medium and low consumption tiers (>5 DN, <5DN and <1 DN respectively). ARPU estimates are adapted from GSMA (GSMA, 2020) for Côte D'Ivoire, Mali, Senegal and Kenya (high: $8, medium: $6, low: $2), and for Tanzania and Uganda (high: $8, medium: $3, low: $2). Revenue is converted to the Net Present Value (NPV) over the assessment period using a discount rate of 5%.

### 3.3. High-resolution spatial estimation of least-cost networks

The network design module estimates the least-cost design to connect communities without cellular coverage. Firstly, a baseline is established of existing infrastructure using a range of data sources. Long distance fiber links are extracted from the African Terrestrial Fiber map (Network Startup Resource Center, 2020), and fiber POPs are estimated based on large settlements exceeding 20,000 inhabitants located within 5 km of a fiber edge. Secondly, existing sites in each region are estimated to obtain the total existing site density using either geolocated site data, or disaggregated estimates of tower counts by country. In the case of the latter, equation (3) details how the sites ($Sites_i$) in the $i$th area are rank estimated given the local population ($Pop_i$), the total number of sites nationally ($Total\_Sites$), the total population nationally ($Total\_Pop$), and the percentage of the population covered nationally with cell phone access ($Total\_Coverage$).

$$Sites_i = Pop_i \cdot \frac{Total\_Sites}{(Total\_Pop \cdot (Total\_Coverage/100))} \qquad (3)$$

To allocate these sites, all regions are sorted based on population density, with the highest population density areas at the top of the list, and lowest population density areas at the bottom. The sites are allocated to the most densely populated regions first using equation (3), meaning at some point all towers are allocated and those areas at the bottom of the ranked list receive no existing assets. These remaining areas are therefore the places of existing market failure which need serving.

Actual site data is provided by the governments of Senegal and Kenya (15,302 and 16,985 cells respectively). Tower count estimates are used of 4,412 in Côte d'Ivoire, 1000 in Mali and 3,554 in Uganda, as well as an estimate of 8,287 cells in Tanzania (TowerXchange, 2018). Sites are treated as having a mean number of three cells per site. Of the total existing site density in each area, the hypothetical MNO modeled has a site density relative to its market share. So, an MNO with 30% market share, is treated as having a site density which is approximately 30% of the total site density. Sites are allocated a technology, such as 2G, 3G or 4G, by intersecting the estimated site locations with the coverage map polygons from the global Mobile Coverage Explorer (Collins Bartholomew, 2019).

To estimate baseline capacity both current and future spectrum bands are identified over the assessment period. Average downlink spectrum portfolios for a hypothetical MNO are identified for Côte D'Ivoire (3G: 15MHz@2100MHz and 4G: 10MHz@ 800MHz), Mali (3G: 10MHz@2100MHz and 4G: 10MHz@700MHz), Senegal (3G: 10MHz@1800MHz and 2100MHz, and 4G 10MHz@800MHz and 1800MHz), Kenya (3G: 20MHz@1800MHz and 4G: 10MHz@700MHz and 800MHz), Tanzania (3G: 10MHz@1800 MHz and 2100MHz and 4G: 10MHz@ 700MHz and 1800MHz) and Uganda (3G: 10MHz@1800 MHz and 2100MHz, and 4G: 10MHz@800MHz and 1800 MHz).

From the literature, a method is used to estimate downlink network capacity based on spectral efficiency, the site density and spectrum bandwidth. The open-source python simulator can estimate cellular capacity for 3G, 4G and 5G using a 3GPP stochastic propagation model (ETSI TR 138 901) to simulate the path loss attributable to irregular terrain, buildings and other environmental cluster for different radio frequencies. A transmitter height of 30m is used along with a power of 40 dBm, with all detailed simulation parameters reported in the original publications (Oughton et al., 2019). Both 3G HSPA+ and 4G LTE use 2x2 MIMO up to 64 QAM. Spectral efficiency values for different technologies are mapped to the Signal-to-Interference-plus-Noise (SINR) ratio using either 3G or 4G modulation and coding lookup tables, using a standard cellular dimensioning approach (Holma and Toskala, 2011). Each macro site has three sectors, following a standard cellular network dimensioning method, hence leading to hexagonal cell areas. To obtain the least-cost RAN design for a specific traffic demand, the site density is minimized. After subtracting existing sites from the minimum number of total sites, the estimated quantity of required greenfield or upgraded brownfield sites can be estimated. A set of capacity-demand lookup tables are then generated for the model which enables site density to be mapped to a mean spectral efficiency, for each generation, frequency band and environment (urban or rural), as illustrated in Figure 4.

Figure 4 Stochastic simulation results by frequency and technology

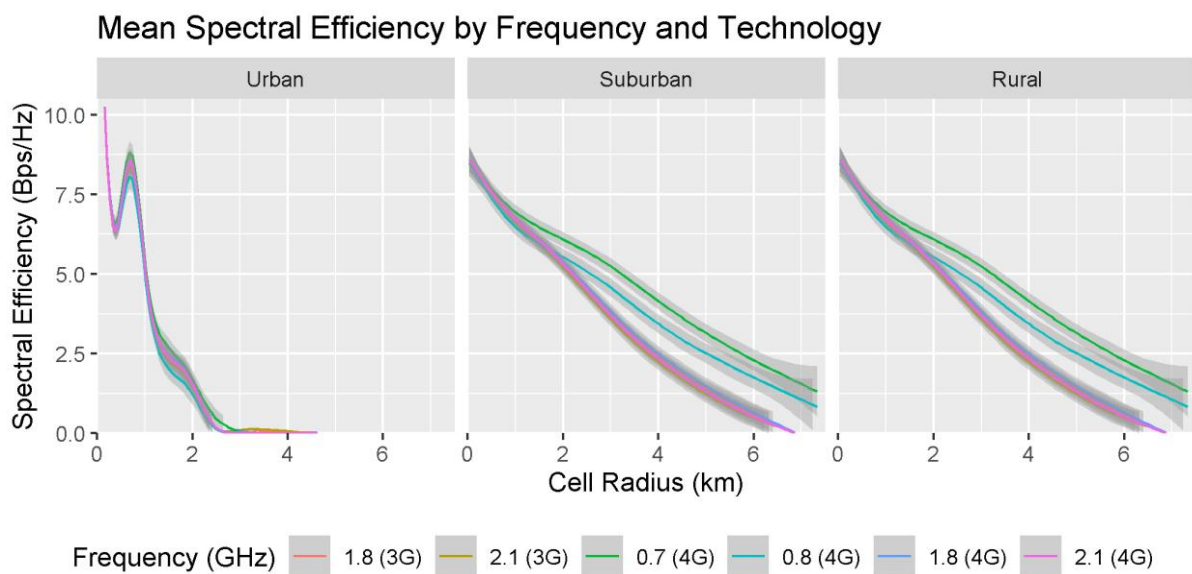

The backhaul for the sites in each region is estimated based on statistics reported by GSMA on the existing composition of technology types by global region (GSMA, 2019b). For Sub-Saharan Africa the current backhaul composition is 4% fixed fiber, 6% fixed copper, 84% wireless and the remaining 6% using satellite. A least-cost design is also used to connect areas via a backhaul link into the main fiber network. Using a minimum spanning tree the cheapest network structure to connect all regional nodes and sites is estimated, which can either be linked using fiber or a wireless technologies, as illustrated in Figure 5. The existing fiber network is in black, while new core fiber links are in orange, and red links are either fiber or multi-hop wireless connections depending on the strategy.

Figure 5 Example of a least-cost network design for Kenya

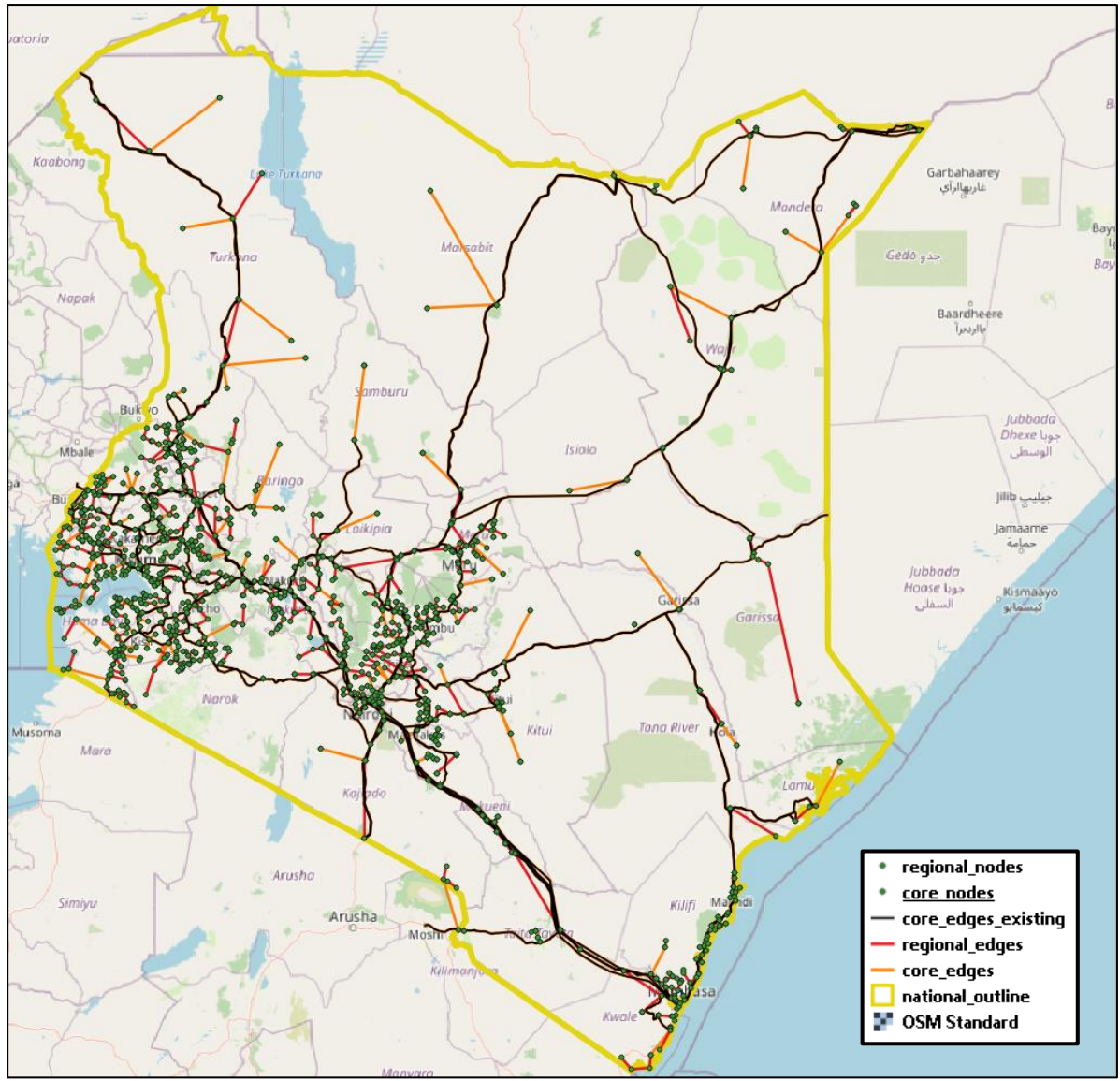

Once the additional sites required and the distance of the backhaul links are known, costs can be developed using mean estimates from a literature survey (5G NORMA, 2016; Frias and Pérez, 2012; Johansson et al., 2004; Markendahl and Mäkitalo, 2010; Oughton et al., 2021; Paolini and Fili, 2012; Smail and Weijia, 2017). Per site capex costs include $39k for all active equipment, $47k to build a full 30m tower and $27k for installation. Per site opex costs include operation and maintenance of $7.4k, power of $2.2k, along with site rental of $15k (urban), $9.9k (suburban) and $2k (rural). For the backhaul, fiber costs per meter are $25, $15 and $10 for urban, suburban and rural respectively. Wireless backhaul costs are based on $10k, $20k and $40k for each small (<10km), medium (<20km) and large (<40km) backhaul unit (of which two are required to form a wireless connection). Connections over 40km require multiple hops. Core and backhaul links use an annual opex of 10% of the initial capex required for all active equipment to cover energy, maintenance and operation. An administration cost of 10% of the RAN cost is added to cover all necessary activities including subscriber acquisition, marketing and R&D. The cost calculations estimate the NPV over the assessment period (2020-2030) using a discount rate of 5%. A market-set Weighted Average Cost of Capital of 15% is used reflecting the risk of capital lending. The network architectures for each type of cell site are shown in Figure 6, with only minor differences between 2G, 3G and 4G.

Figure 6 Cell site design for different technologies

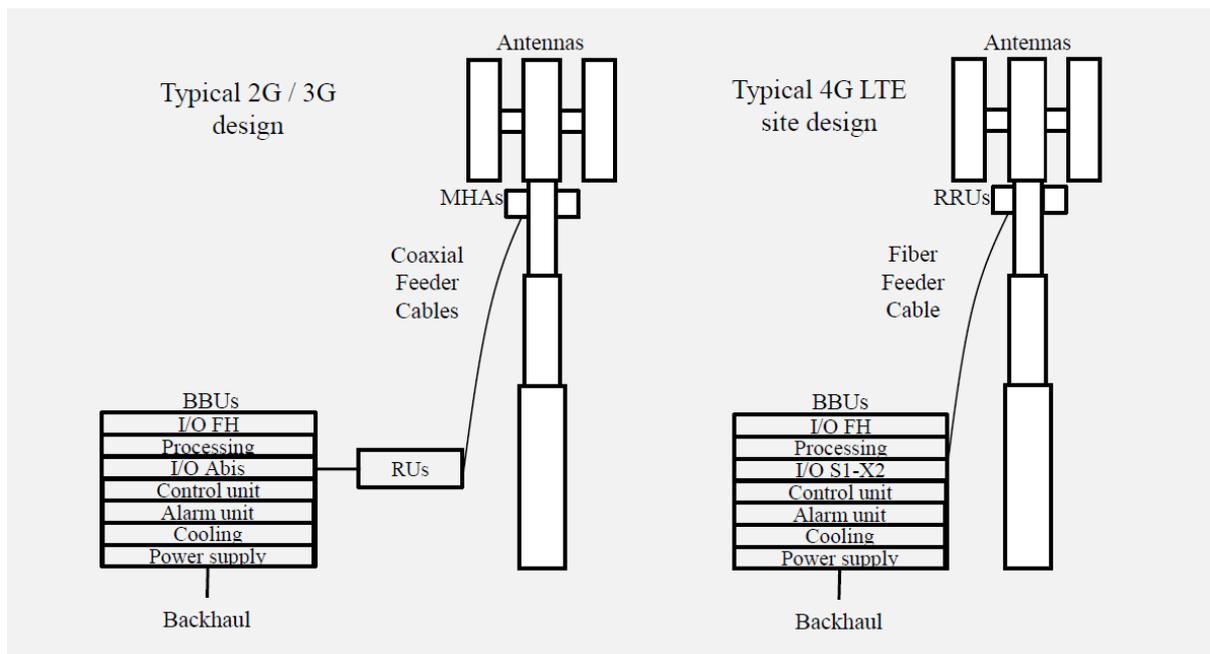

## 3.4. Assessment method

In the assessment module a subscriber acquisition cost is estimated for each country and added to the roll-out cost for each region for coverage and capacity spectrum (<1GHz and >1GHz respectively). Spectrum costs for the $i$th region are also estimated for all frequencies as follows: $Spectrum_i = \sum f \left(Price_{MHz} \cdot BW_f \cdot Pop_i\right)$. Spectrum prices broadly reflect historical costs in Côte D'Ivoire, Mali and Uganda (coverage: $0.02/MHz/population and capacity: $0.01/MHz/population), and Senegal, Kenya and Tanzania (coverage: $0.1/MHz/population and capacity: $0.08/MHz/population).

The taxation rate is treated as 30% of the network investment, and the MNO is allocated a 10% profit margin after all other costs are accounted for as the return for taking the investment risk associated with the network deployment. If gross profits are extracted, infrastructure can only be viably deployed in urban and suburban areas, leaving large rural areas uncovered, which is not conducive for broadband universal serve. Hence, excess capital beyond the 10% profit margin is reallocated to the next most viable region via a process of user cross-subsidization. This is essentially a universal service obligation. After this reallocation process is completed, any areas which remain unviable will require state funded subsidization, but as state funds are limited, this is therefore a last resort.

Additionally, the private cost to the MNO in the $i$th region is estimated based on the sum of the network, admin and operations, spectrum, taxes and profit ($PrivateCost_i = Network_i + Administration_i + Spectrum_i + Taxes_i + Profit_i$). Finally, the net cost to government in the $i$th region is treated as the required state subsidy minus any revenues gained from spectrum fees and taxation ($GovernmentCost_i = Subsidy_i - (Spectrum_i + Tax_i)$).

## 4. Results

The cumulative private cost by population decile is presented in Figure 7 for each of the six countries analyzed under the low, baseline and high adoption scenarios. The results show that 3G is more costly than 4G to deploy, as the lower spectral efficiency means more sites are needed to provide broadband universal service of 10 Mbps. On average 3G is between 11-51% more expensive than 4G when using either a wireless or fiber backhaul, suggesting there is motive for 'leapfrogging' straight to a more recent cellular technology. Add to this the ability to have an internet protocol-based core network (Evolved Packet Core) and 4G becomes an even more appealing technology choice. Additionally, fiber is generally twice the cost of using a wireless backhaul for delivering broadband universal service.

Figure 7 Performance of technology strategies

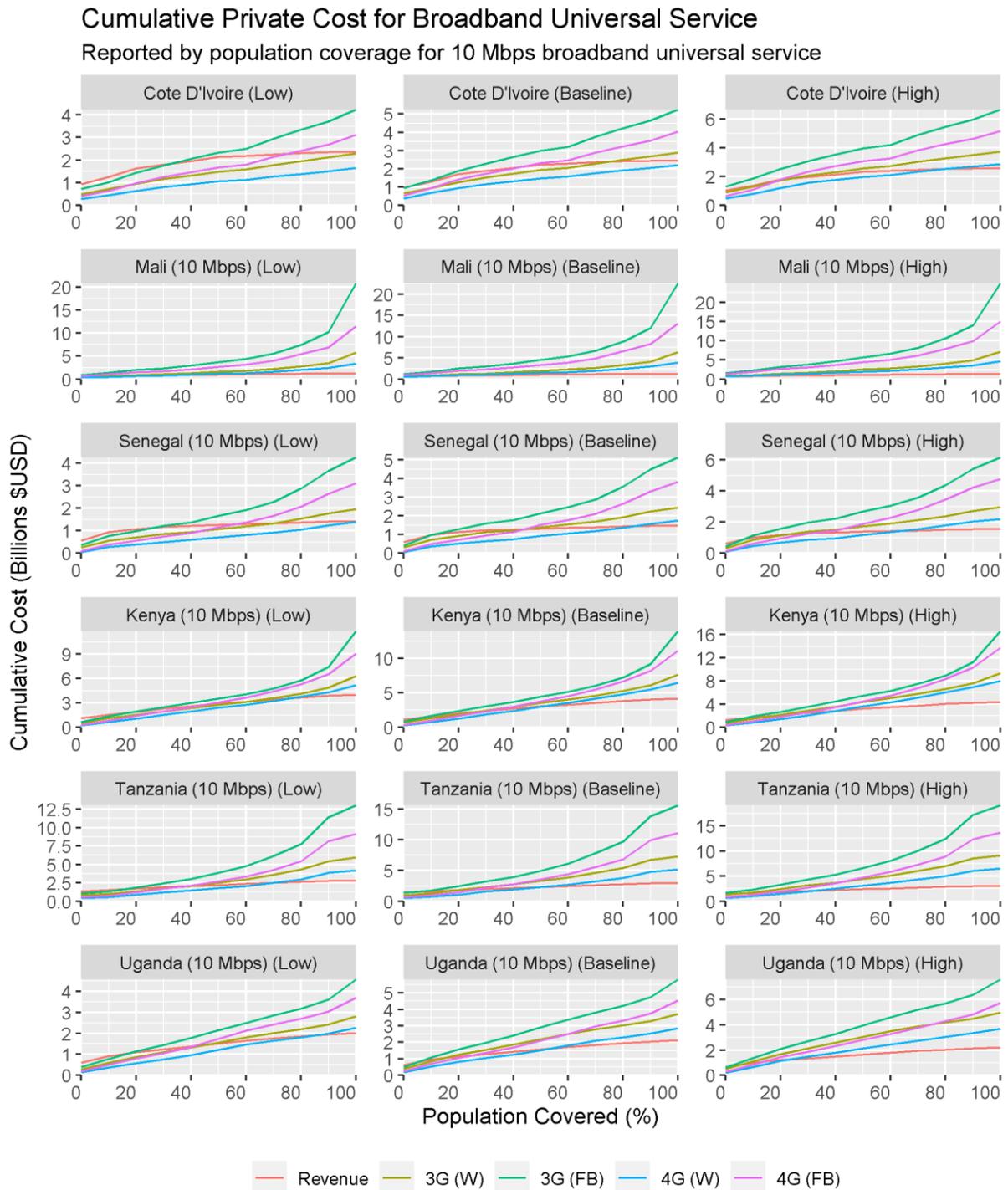

Although revenue is very much dependent on market structure and the degree of competition, a cumulative private revenue curve for the hypothetical MNO modeled is provided to help guide the plausibility of different strategies in delivering broadband universal service. The relationship between plausible revenue and cost is critical for establishing the quantity of user cross-subsidization and required state subsidy. Only a small uplift in revenue is achieved as the scenario adoption rate increases, as most unconnected users are currently in rural areas, where the ARPU is very low. The plots demonstrate most revenue is generated from the first 50% of the densest population deciles, while the remaining deciles general little revenue and are simultaneously the hardest-to-reach making infrastructure deployment very challenging, exacerbating digital inequality.

Figure 8 Cost profile for technology strategies to achieve broadband universal service

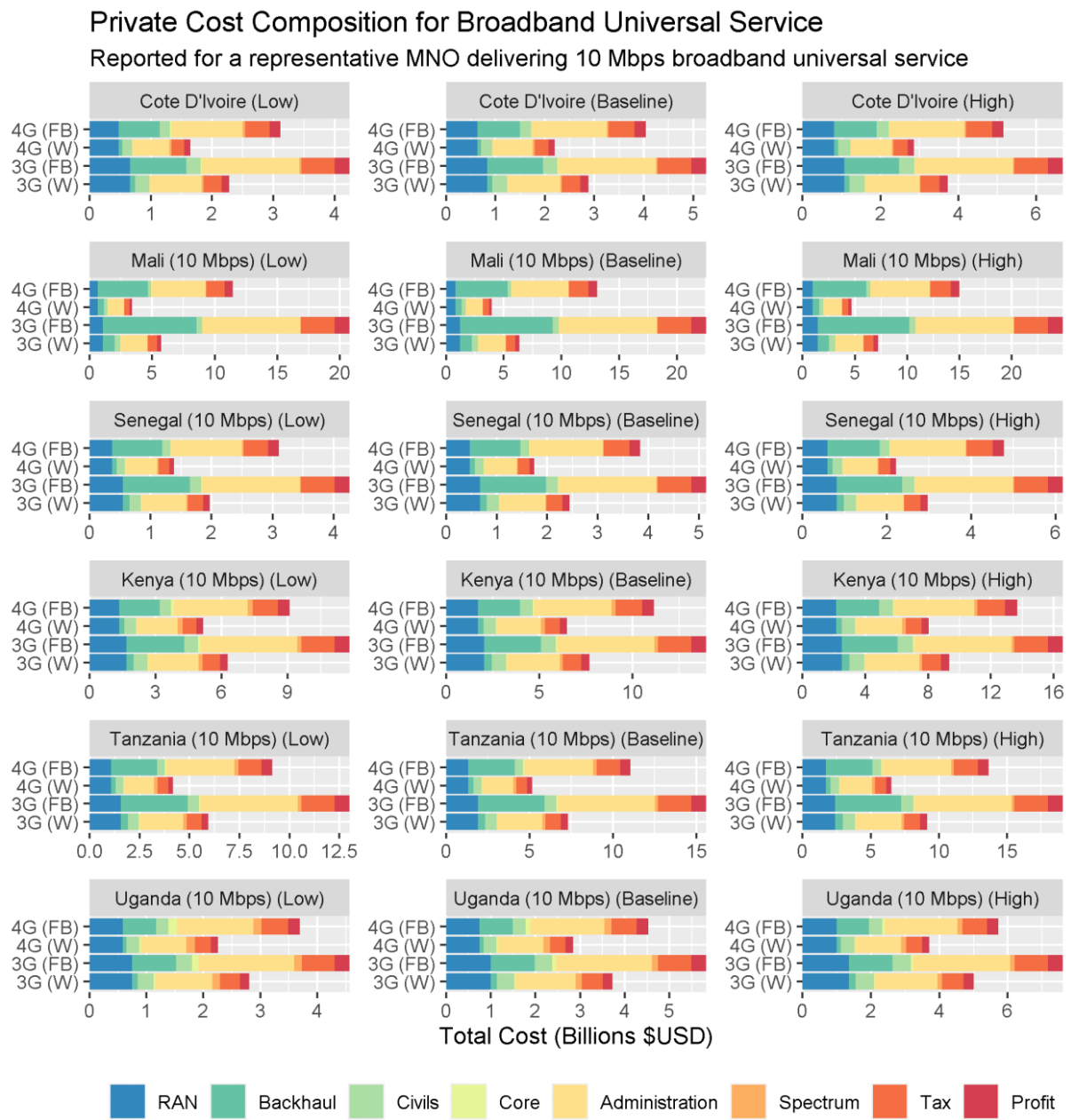

Fortunately, the results suggest that full 4G coverage using a wireless backhaul (with 10 Mbps per user) is potentially within reach for certain countries over the next decade. While this varies by scenario, Cote d'Ivoire could plausibly deliver broadband universal service of 10 Mbps to 100% population coverage in the baseline, while Senegal could achieve 85% and Uganda 55% population coverage.

The composition of the private cost is visualized for each technology in Figure 8, demonstrating the structure of the investment required to achieve comprehensive broadband universal service. Firstly, the magnitude of the overall cost reemphasizes the key messages from Figure 7 in that 4G (W) is the most cost-efficient technology. Generally, the cost composition varies considerably depending on the context, so countries with large remote areas have much larger backhaul costs (e.g. Mali and Tanzania). The total network cost (capex and opex) is approximately ~42% of the required investment, with the other ~58% of the cost composition comprised of administration, spectrum, taxation etc.

Figure 9 Required government investment for broadband universal service

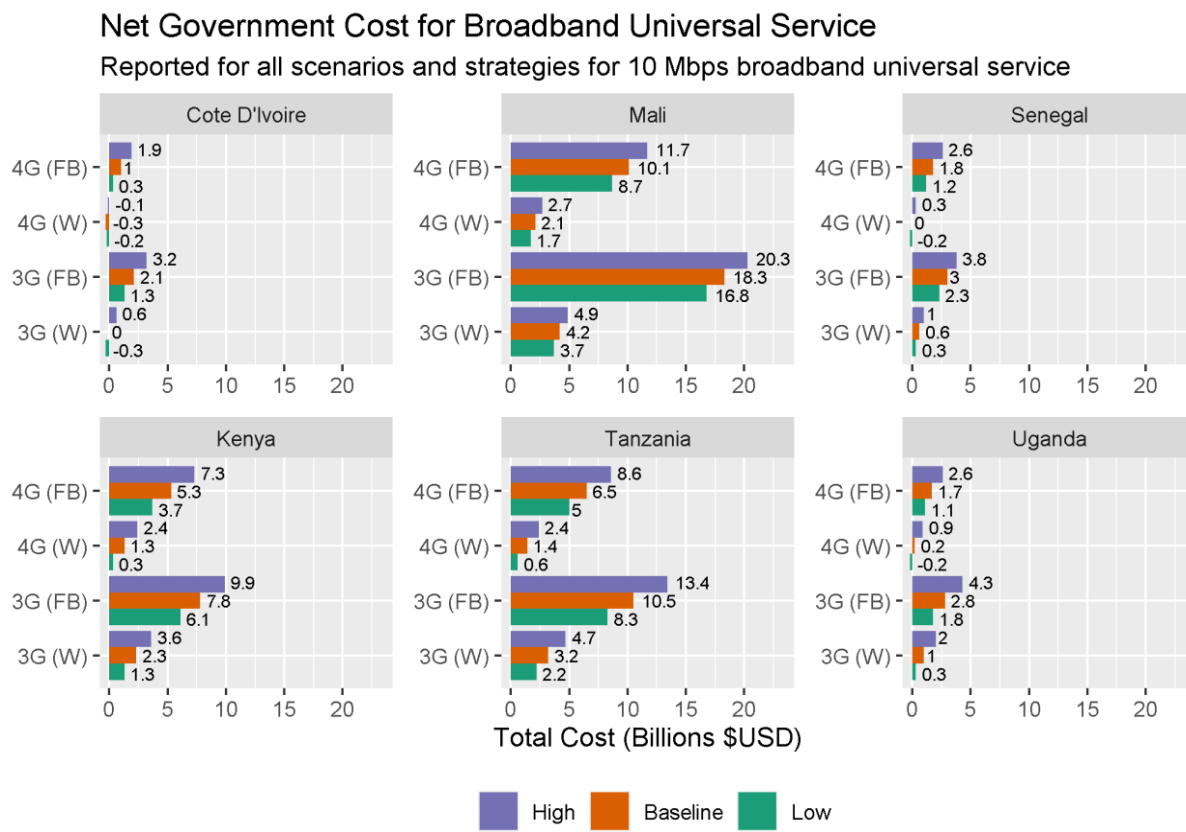

In most cases the potential revenue to support universal broadband falls short of the actual required network investment and operation over the study period. Figure 9 reports the net government cost which represents the necessary infrastructure subsidy required after gains from spectrum and tax revenues. Positive values represent the cost to government, whereas negative values represent a net revenue to government.

By using a more expensive technology, such as 3G, the magnitude of the government cost in the form of an infrastructure subsidy increases in order to achieve broadband universal service. For example, with 4G (W) 15% of the total cost can be paid via a user cross-subsidy, resulting in 19% of the remaining cost shortfall coming from a state subsidy. In comparison, with 3G (W) only 7% of the total cost can come from a user cross-subsidy, requiring a much larger state subsidy representing 29% of the cost. Importantly, with only certain scenarios and strategies exhibiting negative values, therefore providing a net gain to government, Figure 9 establishes that extracting spectrum and taxation revenues in unviable markets provides no net benefit. To achieve broadband universal service, for every $1 of revenue taken by government, this equates to $1 of expenditure in the form of an infrastructure subsidy in unviable areas.

Given these large costs, Figure 10 illustrates the cost savings possible from infrastructure sharing strategies. Passive site and backhaul sharing strategies exhibit substantial cost savings around 33% and 55% respectively. Moreover, a Shared Rural Network provides impressive efficiencies given the approach preserves infrastructure competitive in viable areas, with a 78% saving in the baseline. In contrast, active sharing options via a Multi Operator Radio Access Network (MORAN) saw a very high saving of 260% in the baseline but comes with the caveat that such a strategy would sacrifice competitive infrastructure effects. Such efficiencies result in the cumulative private cost curve lowering significantly in Figure 10 below the potential cumulative revenue, indicating much improved viability for broadband universal service.

Figure 10 Performance of business model infrastructure sharing strategies

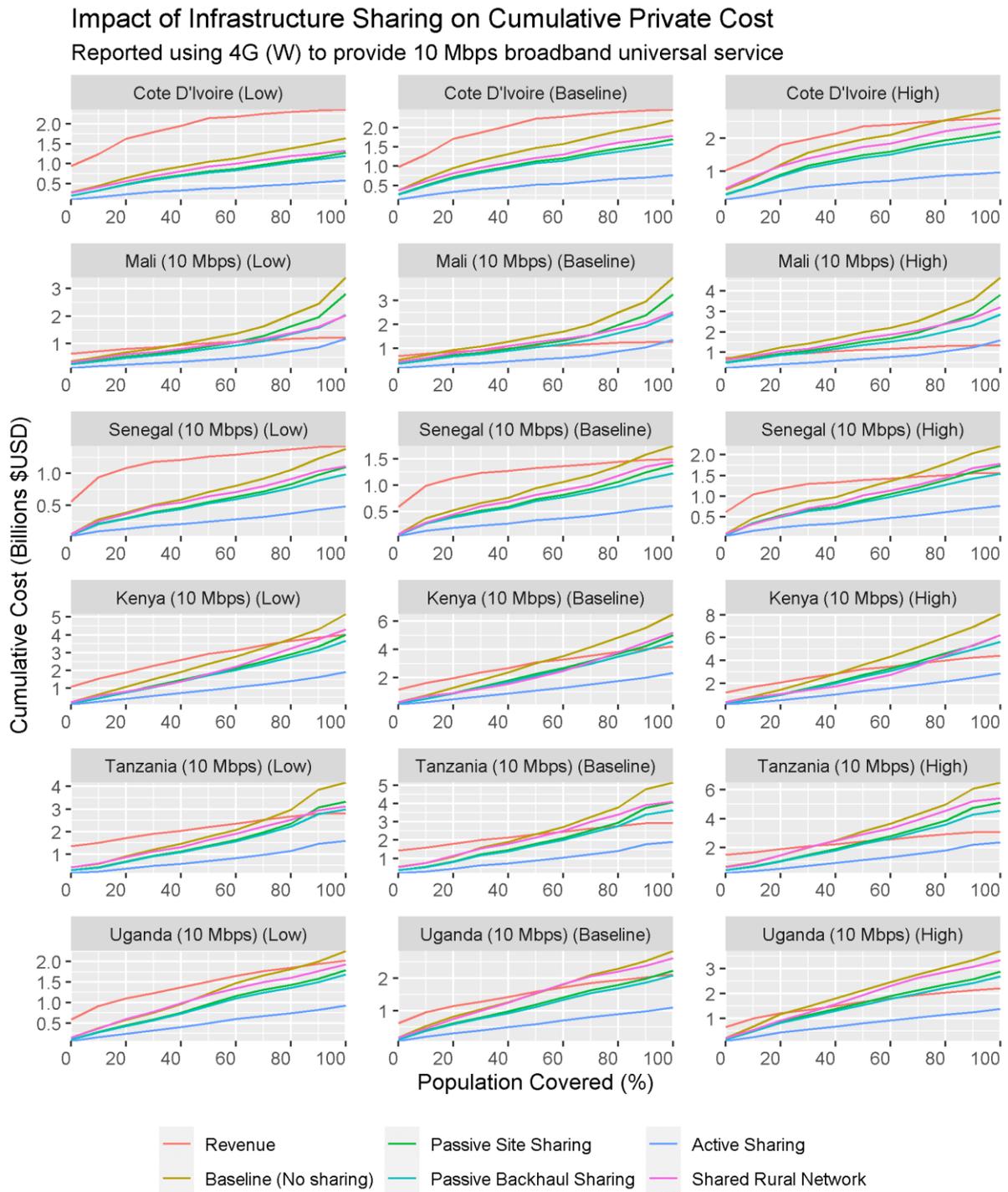

# 5. Discussion and Conclusion

Having reported the results, the research questions will now be discussed, with the first being as follows:

*Which technologies should governments encourage to enable broadband universal service?*

Out of the four different technology options tested the cheapest deployment option across all scenarios and strategies was the combination of 4G with a wireless backhaul. This provides substantial evidence for those countries currently aiming to cycle through the cellular generations sequentially, to instead 'leapfrog' to providing 4G in underserved areas. Most of this cost saving comes from operating fewer sites, thanks to 4G being a more spectrally efficient technology, but there are also numerous benefit from having a more flexible IP-enabled core network making the 4G case even more convincing. The analysis demonstrates that the technology strategy has a very large impact on the viability of universal service broadband, and that using 4G (W) is the most cost-efficient way to push coverage out to rural and remote areas with low ARPU.

Fortunately, under the baseline Cote d'Ivoire could plausibly deliver broadband universal service of 10 Mbps, whereas Senegal and Uganda could achieve up to 85% and 55% of population coverage respectively, without additional policy measures. But governments should be aware that by deploying a higher specification technology (e.g. 4G with fiber) to more dense population deciles, there is less capital available to redeploy to unviable areas via user cross-subsidy, increasing the required state subsidy. Indeed, 4G (W) allowed a user cross-subsidy of 15% of the total cost requiring only 19% of the cost shortfall to be provided via a state subsidy. Whereas with 3G (W), a user cross-subsidy of only 7% of the total cost was available, meaning governments must contribute the remaining 29%. Governments should ensure the regulatory environment encourages a degree of user cross-subsidization from viable to unviable areas, in order to provide broadband universal service.

Importantly, with only certain scenarios and strategies exhibiting a net gain to government, Figure 9 suggests that extracting spectrum and taxation revenues in unviable markets provides no net benefit. To achieve broadband universal service, if a government extracts $1 of revenue from private operators, this consequently equates to $1 of government expenditure in the form of an infrastructure subsidy to unviable areas. Having evaluated the first question, the second will now be discussed:

*What level of infrastructure sharing should governments encourage to help deliver broadband universal service?*

Infrastructure sharing has a very large impact on the cost of delivery, especially in hard-to-reach areas. The caveat to these results is that governments must balance the desire to push out service to unviable areas by reducing supply-side costs against the benefits of competitive infrastructure markets. From decades of economic research, dynamic competition has demonstrated positive outcomes for consumers and the wider macroeconomy. However, issues arise in areas of market failure where the costs of supply exceed the potential revenues obtained from the available demand, therefore giving rise to the phenomenon known as the 'digital divide'.

The concern with the strategies tested, particularly active infrastructure sharing, is that infrastructure consolidation could decrease the level of market competition, which could be an unwise path to take. In contrast, strategies which provided on average 33% and 55% cost saving for site sharing and backhaul sharing respectively could prove promising, but again are nationally homogenous in their approach, with no differentiation between sharing in viable and unviable areas. Therefore, the most promising option is a Shared Rural Network because this approach is capable of balancing competitive markets in viable areas with enhanced sharing in unviable areas. Indeed, up to 78% of the cost saving was achieved by only sharing infrastructure in rural areas, allowing urban and suburban areas to enjoy the benefits of dynamic competition between MNOs. Governments should therefore undertake their own detailed assessments of infrastructure sharing areas deemed to be too unviable to cover with existing terrestrial cellular business models.

Finally, with the UN Broadband Commission exploring a highly ambitious target of 10 Mbps per user, future research should consider the implications of more realistic targets, particularly 2 and 5 Mbps per user, which could be more appropriate for the challenging markets assessed in this paper.